\documentclass[manuscript]{aastex}

\shorttitle{ }

\shortauthors{Cao et al.}

\begin{document}

\title{The hadronic origin of hard gamma-ray spectrum from blazar 1ES 1101-232 }
\author{Gang Cao\altaffilmark{1,2,3} and  Jiancheng Wang\altaffilmark{1,2} }
\email{gcao@ynao.ac.cn,jcwang@ynao.ac.cn}
\altaffiltext{1}{Yunnan Observatories, Chinese Academy of Sciences, Key
Laboratory for the Structure and Evolution of Celestial Objects,
Chinese Academy of Sciences, P.O. Box 110 Kunming, Yunnan Province
650011, P.R. China}
\altaffiltext{2}{Key Laboratory for the Structure and Evolution of Celestial Objects, Chinese Academy
of Sciences, Kunming, Yunnan Province 650011, China}
\altaffiltext{3}{University of Chinese Academy of Sciences, Beijing, China}

\begin{abstract}
The very hard $\gamma$-ray spectrum from distant blazars challenges the traditional synchrotron self-Compton (SSC) model, which may indicate that
there is the contribution of an additional high-energy component beyond the SSC emission. In this paper, we study the possible origin of the hard $\gamma$-ray spectrum from distant blazars. We develop a model to explain the hard $\gamma$-ray spectrum from blazar 1ES 1101-232.
In the model, the optical and X-ray radiation would come from the synchrotron radiation of primary electrons and secondary pairs, the GeV emission would be produced by the SSC process, however, the hard $\gamma$-ray spectrum would originate from the decay of neutral pion produced through proton-photon interactions with the synchrotron radiation photons within the jet.
Our model can explain the observed SED of 1ES 1101-232 well, especially the very hard $\gamma$-ray spectrum.
However, our model requires the very large proton power to efficiently produce the $\gamma$-ray through proton-photon interactions.

\end{abstract}


\section{INTRODUCTION}
Blazars are a special subclass of active galactic nuclei with a relativistic jet oriented at a small angle with respect to the line of sight.
They usually show extreme variability, a high degree of polarization, and a strong non-thermal continuum in the optical/UV band.
Blazars can be divided into two subclasses: flat spectrum radio quasars (FSRQs) and BL Lac objects.
BL Lacs have the lack or weakness of the emission lines, while FSRQs usually present strong broad emission lines.
The spectrum energy distributions (SEDs) of blazars are dominated by non-thermal emission and consist of two distinct, broad components: a low-energy component from radio through UV or X-ray, and a high-energy component from  X-ray to $\gamma$-ray.
It is widely believed  that the low-energy component of blazar SEDs is synchrotron emission from relativistic electrons in the jet, whereas the origin of high component is still a matter of debate.
There are two classes of models to explain high energy emission: leptonic model and hadronic model. In the leptonic model, the high energy emission is produced by inverse Compton (IC) scattering of electrons on a photon field (e.g., B\"{o}ttcher 2007). Possible soft photons are the synchrotron photons produced within the jet (the synchrotron self-Compton, SSC, process; Maraschi et al. 1992; Bloom and Marscher 1996; Zhang et al. 2012 ) or the photons external to the jet (the  External Compton, EC, process; Dermer and Schlickeiser 1993; Sikora et al. 1994; Bl\'azejowski et al. 2000) such as the accretion disk, the broad line region, or the external torus. In the hadronic model, the high energy emission is produced by proton synchrotron or secondary emission from proton-protons and proton-photon interactions \citep{man93,pho00,muc00}.

To data, More than 40 blazars have been detected in the TeV band, most of these objects are high-frequency-peaked BL Lac objects (HBLs).
The primary TeV photons from  distant blazars are absorbed due to pair production by interacting with the extragalactic background light (EBL) photons
\citep{gou66}, the intrinsic TeV spectrum is harder than observed one. The TeV emission from HBLs can generally explained by the standard SSC model
well. However, the observations of blazars with very hard gamma-ray spectrum from some HBLs presented a challenge to the standard SSC model.
One characteristic case is  the very high energy (VHE) $\gamma$-ray emission from distant blazar 1ES 1101-232, which is detected by the
H.E.S.S. array of Cherenkov telescopes \citep{aha06,aha07a}.
The VHE $\gamma$-ray data implied a very hard intrinsic spectrum with a peak in the SED above 3 TeV and a photon index $\Gamma_{\rm int}\leq1.5$,
even when corrected by absorption for the lowest EBL level \citep{aha07a}.
A similar behavior has also been detected in the TeV blazar 1ES 0229+200 \citep{aha07b}, Mrk 501\citep{ner12} and 1ES 0414+009 \citep{abr12}.
It is very difficult to produce such hard $\gamma$-ray spectrum in the context of the one-zone SSC scenario.
The standard shock acceleration theory predicted an electron spectrum with index $p \geq 2$, which corresponds to an intrinsic photon index $\Gamma_{\rm int}\geq1.5$. On the other hand, The suppression of scattering cross section at high energies due to  Klein-Nishina (KN) effect would make the $\gamma$-ray spectrum to be steeper.

Recently, \citet{sah13a} studied the orphan TeV flare of 1ES 1959+650 based on hadronic model, they explained this orphan TeV flare  as the decay of neutral pions produced through the p$\gamma$ interactions with the low-energy tails of SSC photons.
The decay of neutral pions are also applied to explain the multi-TeV $\gamma$-ray emission from the Centaurus A and M 87 \citep{sah12,sah13b}.
\citet{mas13} studied the X-ray and $\gamma$-ray variability of Mrk 421 using leptohadronic model, they suggested that the model based on pion decay
can produce the quadratic variability between the X-ray and TeV band in a much more natural way.
Moreover, the H.E.S.S. observation of 1ES 1101-232 shows no evidence for significant variability on any time scale.
The hard $\gamma$-ray spectrum  may indicate that there is the contribution of an additional spectral component beyond the common SSC emission.
It is possible that hard $\gamma$-ray spectrum has a hadronic origin.
Motivated by the above arguments, in this paper we study the  possible origin of  hard spectrum for distant TeV blazar.
We suggest that the hard $\gamma$-ray spectrum for blazar 1ES 1101-232 may originate from the decay of neutral pion produced through proton-photon interactions with the synchrotron photons within the jet.

In Section 2 we give a brief description of the model. In Section 3 we apply the model to explain the very hard TeV spectrum from 1ES 1101-232. The discussion and conclusion are presented in Section 4.  Throughout this paper, we adopt the cosmological parameters of $H_0 = 70 $  km  s$^{-1}$  Mpc$^{-1}$, $\Omega_M = 0.3$, $\Omega_{\Lambda} = 0.7$.

\section{MODEL}

\subsection{SSC Emision}
The SSC model has been successfully used to explain the multi-band SEDs of blazars. We use the model given by \citet{fin08} to calculate synchrotron and SSC flux.
We assume a spherical blob of the jet moving with Lorentz factor $\Gamma$ at a small angle $\theta$ with the line of sight, which is filled by relativistic electrons with a uniform magnetic field.
 The observed radiation is strongly boosted by a Doppler factor $\delta_{\rm D}=[\Gamma(1-\beta \cos\theta)]^{-1}$.
Since blazar jets are almost aligned to the line of sight of the observer, we assume the Doppler factor $\delta_{\rm D}\simeq\Gamma$.
Throughout the paper, unprimed quantities refer to the observer's frame and  primed quantities to the blob's frame.

We adopt a broken power-law function with a sharp cut-off to describe the electron energy distribution in the emission region:
\begin{equation}
N'_e(\gamma')=\left\{\begin{array}{ll}
K_e\gamma'^{-n_{1}}                                  &\mbox{$\gamma'\leq\gamma'_{\rm b}$},\\
K_e{\gamma'_{\rm b}}^{n_{2}-n_{1}}\gamma'^{-n_{2}}   &\mbox{$\gamma'>\gamma'_{\rm b}$},   \\
\end{array}
\right.
\end{equation}
where $\gamma'_{\rm b}m_ec^2$ is broken energy, $K_e$ is the normalization factor, $n_1$ and $n_2$ are the electron power-law indexes below and above broken energy $\gamma'_{\rm b}m_ec^2$.

\subsection{Secondary Emission From the proton-photon Interactions }
The variable and non-thermal high-energy emission from blazars implied that the jet of blazars can efficiently accelerate particles to very high energy by Fermi shock acceleration.
The electron acceleration is limited by important radiative loss, protons should be accelerated  with higher efficiency and can reach very high energy by the same acceleration mechanisms.
If protons are accelerated to sufficient high energy to reach the threshold for proton-photon interactions,
the high-energy protons would interact with the jet synchrotron photons to produce pions.
The charged pion  would decay to produce the neutrinos and secondary electrons/positrons. The neutrinos from the charged pion
could be detected in neutrino telescopes, such as IceCube \citep{abb11}.
Decay of the neutral pion would result in high-energy $\gamma$-ray emission, which is directly observed by ground-based telescope up to the TeV energy. Also, a correlation between the low-energy and high-energy bump is not necessary. On the other hand,
the decay of neutral pion from proton-photon interactions would describe the high-energy bump of blazar SEDs.

The pion production channels through the $p\gamma$ interactions are
\begin{eqnarray}
p+\gamma\rightarrow\left\{\begin{array}{ll}
p+\pi^{0}                                \\
n+\pi^{+}  \\
\end{array}.
\right.
\end{eqnarray}
The revelent pion decay channels are
\begin{eqnarray}
\pi^{+}&\rightarrow&\mu^{+}+\nu_{\mu}\rightarrow e^{+}+\nu_{e}+\overline{\nu}_{\mu}+\nu_{\mu},\\
\pi^{0}&\rightarrow&\gamma+\gamma.
\end{eqnarray}
The threshold condition for the $p\gamma$ interactions is given by \citep{kel08}
\begin{equation}
2E'_p\epsilon'_{\rm ph}(1-\beta_p\cos\theta) = (2m_{\pi}m_p+m_{\pi}^2)c^4,
\end{equation}
where $E'_p$ and $\epsilon'_{\rm ph}$ are the proton and soft photon energy in the comoving frame.
For high energy protons we set $\beta_p\simeq1$. In the case of a head-on collision ($\theta=\pi$),
Equation (5) implies
\begin{equation}
E'_p\epsilon'_{\rm ph}\simeq 0.07 {\rm GeV}^2
\end{equation}
In the single-pion production channels, each pion carries $\sim0.2$ of proton energy. Consider that each $\pi^{0}$ decay into two $\gamma$-ray photons,
the observed $\gamma$-ray energy from $\pi^{0}$ decay is
\begin{equation}
E_{\gamma}=\frac{1}{10}\frac{\delta_{\rm D}}{1+z}E'_p.
\end{equation}

We follow the method of \citet{ato03} to calculate the spectrum from $\pi^{0}$ decay.
The cooling rate due to the p$\gamma$ interactions with the synchrotron radiation field is given by \citep{ste68}
\begin{equation}
t_{p\gamma,\pi}^{-1}(\gamma'_p)=\frac{c}{2\gamma'{_p^2}}\int_{\frac{\epsilon'_{\rm th}}{2\gamma'_p}}^{\infty}d\epsilon'\frac{n'_{\rm syn}
(\epsilon')}{\epsilon'^2}
\int_{\epsilon'_{\rm th}}^{2\gamma'_p\epsilon'}dE'\sigma_{p\gamma}(E')K_{p\gamma}(E')E',
\end{equation}
where $\epsilon'_{\rm th}$=145 MeV is the photon threshold energy, $n'_{\rm syn}$ is the number density of the synchrotron radiation given by \citet{fin08}.
The section cross $\sigma_{p\gamma}(E')$ is approximately given by \citep{ato03,rey09}
\begin{equation}
\sigma_{p\gamma}(E')=3.4\times10^{-28}\rm cm^2\Theta(E'-200 \rm MeV)\Theta(500 \rm MeV- E')+ 1.2\times10^{-28}\rm cm^2\Theta(E'-500 \rm MeV),
\end{equation}
and the inelasticity coefficient is
\begin{equation}
K_{p\gamma}(E')=0.2\Theta(E'-200 \rm MeV)\Theta(500 \rm MeV- E')+ 0.6\Theta(E'-500 \rm MeV).
\end{equation}
The corresponding collision rate is given by a similar expression
\begin{equation}
\omega_{p\gamma,\pi}(\gamma'_p)=\frac{c}{2\gamma'{_p^2}}\int_{\frac{\epsilon'_{\rm th}}{2\gamma'_p}}^{\infty}d\epsilon'\frac{n'_{\rm syn}
(\epsilon')}{\epsilon'^2}
\int_{\epsilon'_{\rm th}}^{2\gamma'_p\epsilon'}dE'\sigma_{p\gamma}(E')E'.
\end{equation}
The photon emissivity from  $\pi^{0}$ decay in the $\delta$-function approximation is given by \citep{ato03,rom08}
\begin{equation}
j'(E'_{\gamma})
=20N'_p(10E'_{\gamma})\omega_{p\gamma,\pi}(10E'_{\gamma})n'_{\pi^{0}}(10E'_{\gamma}),
\end{equation}
where $n'_{\pi^{0}}=p_1/2+p_2$ is the mean number of neutral pions created per collision,  $p_1$ and $p_2=1-p_1$ are possibilities of the p$\gamma$ interaction in single-pion and multi-pion channels, and $N'_p$ is the  proton energy distribution.  We can define a mean inelasticity as
$\overline{K}_{p\gamma,\pi}=t_{p\gamma,\pi}^{-1}(\gamma'_p)\omega^{-1}_{p\gamma,\pi}(\gamma'_p)$, then the  possibilities $p_{1,2}$ can be calculated from the relation of $p_1=\frac{K_2-\overline{K}_{p\gamma}}{K_2-K_1}$, where $K_1$=0.2 and $K_2$=0.6.
Taking into account the relativistic and cosmological effects,  we get the observed flux from the secondary emission as \citep{lin85}
\begin{equation}
F(E_{\gamma})=\frac{\delta_{\rm D}^3(1+z)V'}{d_{\rm L}^2}E'_{\gamma}j'(E'_{\gamma}),
\end{equation}
where $z$ is the redshift of the source, $d_{\rm L}$ is the  luminosity distance,  $V'$ is the volume of the emission region, and
$E_{\gamma}=\delta_{\rm D}E'_{\gamma}/(1+z)$ is the photon energy in the observed frame.

The emissivity of secondary pairs can be estimated in the same way. Both in the single-pion and multiple-pion channel, each charged pion
has an energy $\sim0.2E_{\rm p}$. This energy is equally distributed among the products of its decay, hence the energy of each electron/positron
is $E_{\rm e}\sim0.05E_{\rm p}$.  The emissivity of secondary pairs is given by \citep{rom08}
\begin{equation}
Q'(E'_{e^{\pm}})
=20N'_p(20E'_{e^{\pm}})\omega_{p\gamma,\pi}(20E'_{e^{\pm}})n'_{\pi^{\pm}}(20E'_{e^{\pm}}),
\end{equation}
where $n'_{\pi^{\pm}}=p_1/2+2p_2$ is the mean number of charged pions created per collision.
If we neglect the electron escape, the steady electron distribution can be calculated as
\begin{equation}
N'(E'_{e^{\pm}})=4\pi|\frac{dE'_{e^{\pm}}}{dt}|^{-1}\int_{E'}^{E'_{\rm max}}dE'Q'(E'_{e^{\pm}}),
\end{equation}
where $\frac{dE'_{e^{\pm}}}{dt}=\frac{4}{3}c\sigma_{\rm T}\gamma_{e^{\pm}}'^2U_{\rm B}$ is the electron energy loss rate,
$U_{\rm B}=\frac{\rm B^2}{8\pi}$ is the magnetic energy density. In the following,
we use the method of \citet{fin08} to calculate the synchrotron radiation from the secondary electrons.

Protons should be co-accelerated together with electron by the Fermi acceleration mechanism, which predicts the particle distribution index
$\alpha\geq$2.
We adopt the proton injection spectrum with exponential cutoff power-law:
\begin{equation}
N'_{\rm p}=N_0{E'_{\rm p}}^{-\alpha}{\rm exp}(-E'_{\rm p}/E'_{\rm c}) \quad\quad   E'_{\rm p,min}\leq E'_{\rm p}\leq E'_{\rm p,max},
\end{equation}
where $N_0$ is the  initial proton distribution. In our calculation, we assume that the proton injection index to be $\alpha=2.2$,
and the minimum energy of the injection proton to be $E'_{\rm p,min}=145$ MeV, which is the threshold energy of proton-photon interactions.

\section{APPLICATION}
1ES 1101-232 resides in an elliptical galaxy at a redshift of $z=0.186$ \citep{rem89}.  The source has been detected by a range of X-ray instruments in both soft and hard X-ray band \citep{wol00,rei08}. It has been classified as a high-frequency-peaked BL Lac object due to synchrotron peak in the X-ray band \citep{don01}.
\citet{wol00} and \citet{cos02} predicted that 1ES 1101-232
is a very promising candidate for TeV detection. Motivated by these predictions, \citet{aha07a} performed the coordinated optical-X-ray-VHE $\gamma$-ray observation in  2004 and  2005, which results in the discovery of VHE $\gamma$-ray.
Taking the VHE data from H.E.S.S observation together with measurements in the X-ray and optical bands, \citet{aha07a} constructed the truly simultaneous SED of 1ES 1101-232 from the optical to the VHE band  in 2004 and 2005.
However, no simultaneous GeV observation was performed in this multi-wavelength campaign. The GeV data is important to constrain the radiation model of blazars. In the GeV band, EGRET did not detect emission from 1ES 1101-232 \citep{lin96}.
The detections of 1ES 1101-232 were also not reported in the Large Area Telescope (LAT) Bright Active Galactic Nuclei Source List \citep{abd09}.
Fermi-LAT reported the significant detections of 1ES 1101-232 in the First LAT AGN Catalog \citep{abd10a} and in the Second LAT AGN Catalog
\citep{ack11}. However, only the flux upper limits  are available  from the literature \citep{ner10,tav10}.
\citet{fin13} analyzed the Fermi-LAT 3.5 year data collected from 2008 August to 2012 February and reported the GeV spectrum of 1ES 1101-232.
The previous SSC model indicated that the IC peak is expected to be around 100 GeV \citep{wol00,cos02}. However, new observational results  from H.E.S.S  indicate that the source shows a hard VHE spectrum with a peak in the SED above 3 TeV, when corrected by absorption for the lowest EBL level
\citep{aha07a}. In this paper, we use simultaneous SED of 1ES 1101-232 collected by \citet{aha07a} in 2004 and 2005 and the GeV data given by \citet{fin13}.

We use the model described in Section 2 to model the simultaneous SED of 1ES 1101-232 in 2004 June and 2005 March.  The observed data are taken from
\citet{aha07a} and \citet{fin13}. The VHE data is corrected for the EBL absorption considering the low EBL level, as described in \citet{aha07a}. The  modeling parameters are listed in table 1-2.
It can be found that the Poynting flux power $L_{\rm B}$ is slightly larger than the electron power $L_{\rm e}$, the relativistic proton
power $L_{\rm p}$ is far larger than the Poynting flux $L_{\rm B}$, and the total jet power is dominated by relativistic protons. Our model requires the extreme proton power $L_{\rm p}\gtrsim10^{52}$ erg s$^{-1}$ to produce the $\gamma$-ray through the proton-photon interactions efficiently. \citet{bot13} used the leptonic and hadronic model to fit the observed SED of a set of Fermi-LAT-detected blazars, they also showed that the hadronic model usually requires the extreme jet powers, in some case exceeding $10^{49}$ erg s$^{-1}$ in relativistic protons.
The corresponding SED modeling is shown in figure 1-2. It can be seen that the optical and X-ray radiation comes from the synchrotron radiation of primary electrons and secondary pairs, the GeV emission is produced by the SSC process,
the hard TeV emission originates from the $\pi^{0}$ decay produced through proton-photon interactions with the synchrotron photons.
Our model can reproduce the observed SED of 1ES 1101-232 well, particularly  the very hard $\gamma$-ray spectrum.
In our model, the energy of secondary pairs is $E'_{\rm e^{\pm}}\simeq0.05E'_{\rm p}\simeq0.14$ TeV, the corresponding synchrotron frequency in the observed frame is $\nu_{\rm s}=1.4\times10^{18}(\frac{B}{0.4})(\frac{\delta_{\rm D}}{15}$) Hz, which couldn't contribute to the GeV and TeV emission. Moreover, the synchrotron emission from secondary pairs has a small contribution to the optical and X-ray flux compared to that from the primary electrons.

We can estimate the proton energy through the observed VHE $\gamma$-ray energy.
For the source 1ES 1101-232, the highest photon energy is $E_{\gamma}\sim3.7$ TeV. From Equation 7, we can obtain that
the proton energy in the comoving frame is $E'_p\sim2.9$ TeV , which is comparable with the value derived from our SED modeling.
This value corresponding to the observed frame is $E_p\sim37$ TeV, if the proton can escape the source and reach the observer without energy loss. Therefore, the observed proton energy is $E_p<37$ TeV.
The observed $\gamma$-ray energy is in the range $0.36$ TeV$\lesssim E_{\gamma} \lesssim 3.70$ TeV.
Using the Equation 6 and 7, we can obtain that the corresponding target photon energy in proton-photon interactions is
in the range $7.19\times10^{19}$ Hz$\lesssim \nu_{\rm ph} \lesssim 7.31\times10^{20}$ Hz,
which exactly falls in the high-energy part of the synchrotron spectrum, and the SSC flux in this energy range is negligible compared to the synchrotron flux. This indicated that  hard $\gamma$-ray spectrum may come from the $\pi^{0}$ decay produced through proton-photon interactions
with the synchrotron photons.
It should be noted that the  target photon energy of proton-photon interactions lies in the low-energy tails of SSC spectrum in the paper by
\citet{sah12,sah13a,sah13b}, rather than the synchrotron photons.
Moreover, the hard X-ray emission at the energy $\gtrsim$ 100 KeV has been detected by Swift-BAT in several TeV blazars \citep{abd11a,abd11b}.
\citet{kau11} presented the hard X-ray observation of TeV blazar 1ES 0229+200, they showed that the X-ray spectrum is extended up to $\sim$ 100 KeV without any significant cut-off. It is possible that the hard $\gamma$-ray spectrum originates in the $\pi^{0}$ decay
produced through proton-photon interactions with the high-energy parts of the synchrotron photons within the jet.

The efficiency of the proton-photon interaction depends on the density and distribution of the target photons.
In our model, the high-energy proton efficiently interacts with the high-energy parts of the synchrotron photons to product the $\gamma$-ray.
Therefore, the efficiency of the proton-photon interactions depends on the  distribution of the synchrotron photons.
However, the distribution of the synchrotron photons in turn depends on the maximum Lorentz factor $\gamma_{\rm max}$. In figure 3, we show the modeling spectrum with the various values of the maximum Lorentz factor $\gamma'_{\rm max}$. It can be seen that the synchrotron flux at high energies decreases with the decreasing maximum Lorentz factor, which results in the decreasing efficiency of the $\gamma$-ray production through the proton-photon interactions.
Our model requires the maximum Lorentz factor $\gamma'_{\rm max}\geq2\times10^7$ in order to fit the hard $\gamma$-ray spectrum from blazar 1ES 1101-232. The  acceleration timescale of electron is $t_{\rm acc}=\eta\frac{\gamma' m_{\rm e}c}{e\rm B}$. The synchrotron cooling timescale is
$t_{\rm syn}=\frac{6\pi m_{\rm e}c}{\sigma_{\rm T}\gamma'\rm B^2}$. Then the maximum Lorentz factor $\gamma'_{\rm max}$ can be determined by the condition $t_{\rm acc}=t_{\rm syn}$ as
\begin{equation}
\gamma'_{\rm max}\simeq7\times10^7(\frac{\rm B}{0.4})^{-\frac{1}{2}}(\frac{\eta}{10})^{-\frac{1}{2}}.
\end{equation}
Hence, the required maximum Lorentz factor $\gamma'_{\rm max}\geq2\times10^7$ is achievable by the Fermi acceleration mechanism in the jet of blazar 1ES 1101-232.

\section{DISCUSSION AND CONCLUSION}

The proton-proton (pp) interactions with the ambient matter can produce the TeV $\gamma$-ray, as has been proposed to explain the observed SEDs of blazars \citep{dar97,pho00,bea99}. However, the pp interaction usually requires the high plasma density $n_{\rm H}>10^{6}$ cm$^{-3}$ \citep{aha00}, such a high density would not be expected in the environment of BL Lac objects due to small accretion rate. Therefore, the pp interaction is less efficient than the p$\gamma$ interaction.
The different theoretical models has been proposed to explain the hard  $\gamma$-ray spectrum.
\citet{kat06} suggested that the hard $\gamma$-ray spectrum can be explained in the SSC scenarios by assuming a narrow distribution of high-energy electrons with the high low-energy cutoff. It is difficult to maintain such high low-energy cutoff because of the relevant radiation cooling at these
energies. The internal absorption of the TeV $\gamma$-ray spectrum by a very narrow distribution of optical UV photons in the vicinity of AGN can explain the observed hard $\gamma$-ray spectrum \citep{aha08,zac11}. However, there is no indication for the present of such narrow radiation fields in the  vicinity of BL Lac objects. A two-component model has been proposed to explain the observed SED of 1ES 1101-232 \citep{bot08,yan12}. In this model, the low-frequency emission is assumed to originate in the inner region of jet, the hard $\gamma$-ray spectrum can be produced via Compton upscattering of cosmic microwave background (CMB) in an extend jet. However, this scenario requires that the electrons can be accelerated to TeV energy on kiloparsec scales along the jet.
The hard $\gamma$-ray spectrum can be explained by the secondary emission produced along the line of sight by the interactions of the cosmic-ray protons  with the background photons \citep{ess10,ess11}. In this scenario, the low extragalactic magnetic fields (EGMFs) not larger than $10^{-15}$ G is needed to avoid the significant deflection of cosmic-ray protons and secondary electrons, but the strength of EGMF along with the line of sight remains large uncertain. On the other hand, the predicted secondary spectrum fits well the hard TeV spectrum only for the high level EBL model of \citet{ste06}, which are excluded by the Fermi-LAT and H.E.S.S results \citep{abd10b,aha06}. Alternatively, new particles \citep{de07,hor12} and Lorentz invariance violation \citep{pro00} have been invoked to explain the data.

A number of blazars are found to exhibit the correlated variability between the low-energy and VHE band.
Several blazars have  displayed minute timescale variability in the TeV band  \citep{aha07c,alb07}.
No significant correlated variability between the optical/X-ray and $\gamma$-ray bands was observed from this source. At the same time, the VHE data from H.E.S.S show no sign of variability on any time scale.
This seemly supports a hadronic origin of hard $\gamma$-ray spectrum from  blazar 1ES 1101-232. In the paper, we study the possible origin of  hard $\gamma$-ray spectrum. We propose a model to explain the hard $\gamma$-ray spectrum from  blazar 1ES 1101-232.
In this model, the optical and X-ray radiation comes from the synchrotron radiation of primary electrons and secondary pairs, in which the
synchrotron radiation from the secondary pairs has a small contribution to the optical and X-ray flux,
and the GeV emission is attributed to the SSC process, however, the hard $\gamma$-ray spectrum originates in the $\pi^{0}$ decay produced through proton-photon interactions with the synchrotron photons within the jet.
Assuming suitable electron and proton spectra, we obtain a excellent fit to the observed spectrum of  blazar 1ES 1101-232.
The very hard $\gamma$-ray spectrum can be well explained by secondary emission from the $\pi^{0}$ decay in the source.
However, our model requires the extreme proton power in order to efficiently produce the $\gamma$-ray through proton-photon interactions.
Neutrino populations can be expected in the p$\gamma$ interaction, their observation will be the subject of future work.

\begin{acknowledgements}
We thank the anonymous referee for valuable comments and suggestions. We acknowledge financial support from the
National Natural Science Foundation of China 11133006, 11163006, and 11173054, and the Policy Research Program
of the Chinese Academy of Sciences (KJCX2-YW-T24). This work is supported by the Strategic Priority Research Program
\lq\lq The Emergence of Cosmological Structures" of the Chinese Academy of Sciences, grant No. XDB09000000.
\end{acknowledgements}

\clearpage

\begin{figure}
\epsscale{1.}
\plotone{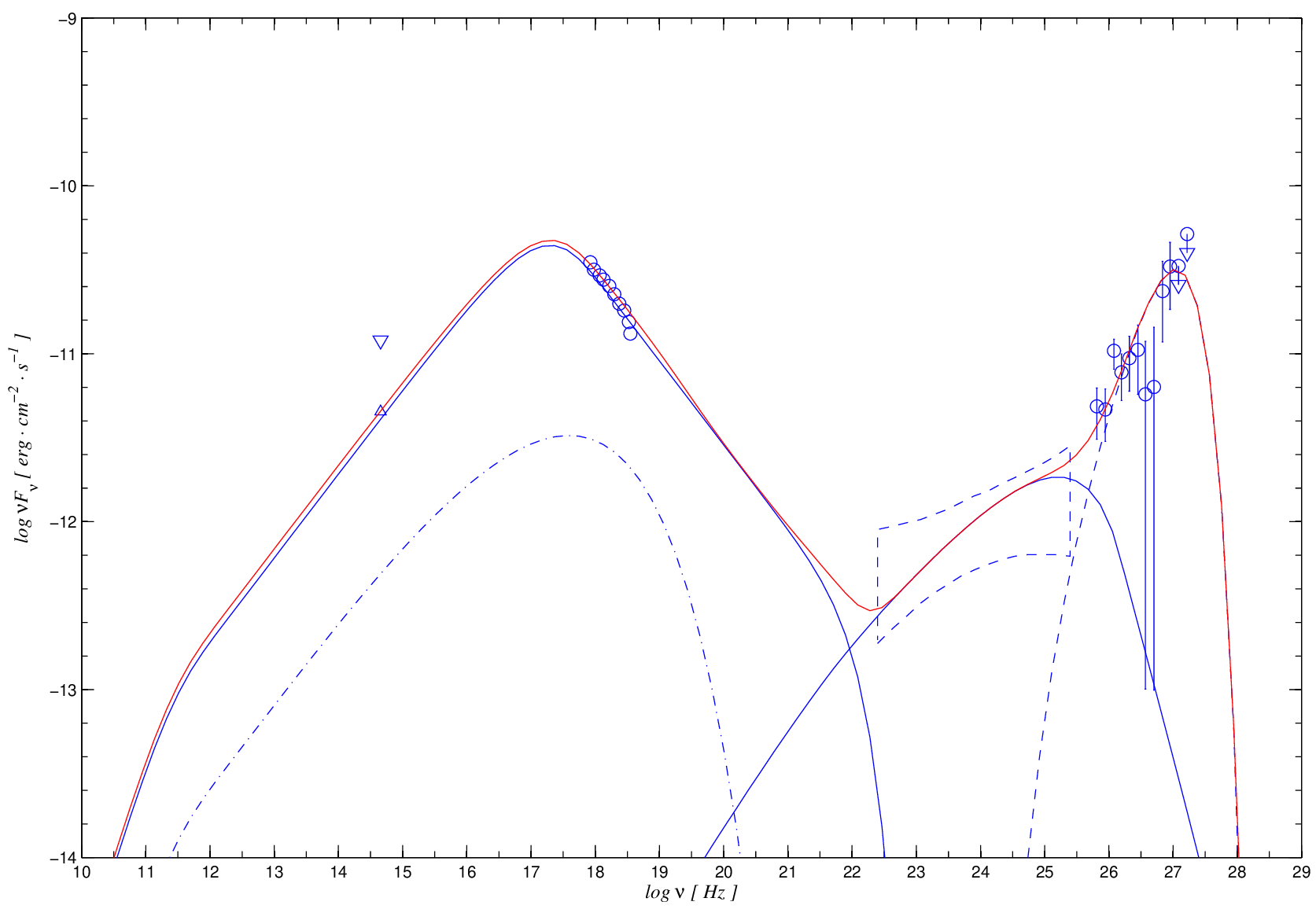}
\caption{The SED of 1ES 1101-232  on 2005 March 5-16 . The blue circles is the simultaneous optical, X-ray and TeV data.
The LAT spectrum is shown as the bowtie.
The blue solid curves represent the synchrotron and  SSC emission, respectively (from left to right).
The blue dot-dashed  curve represents the synchrotron emission from the secondary pairs.
The blue dashed curve represents the secondary emission from the $\pi^{0}$ decay. The red solid curve is the total emission from all spectrum components.
\label{fig1}}
\end{figure}

\begin{figure}
\epsscale{1.}
\plotone{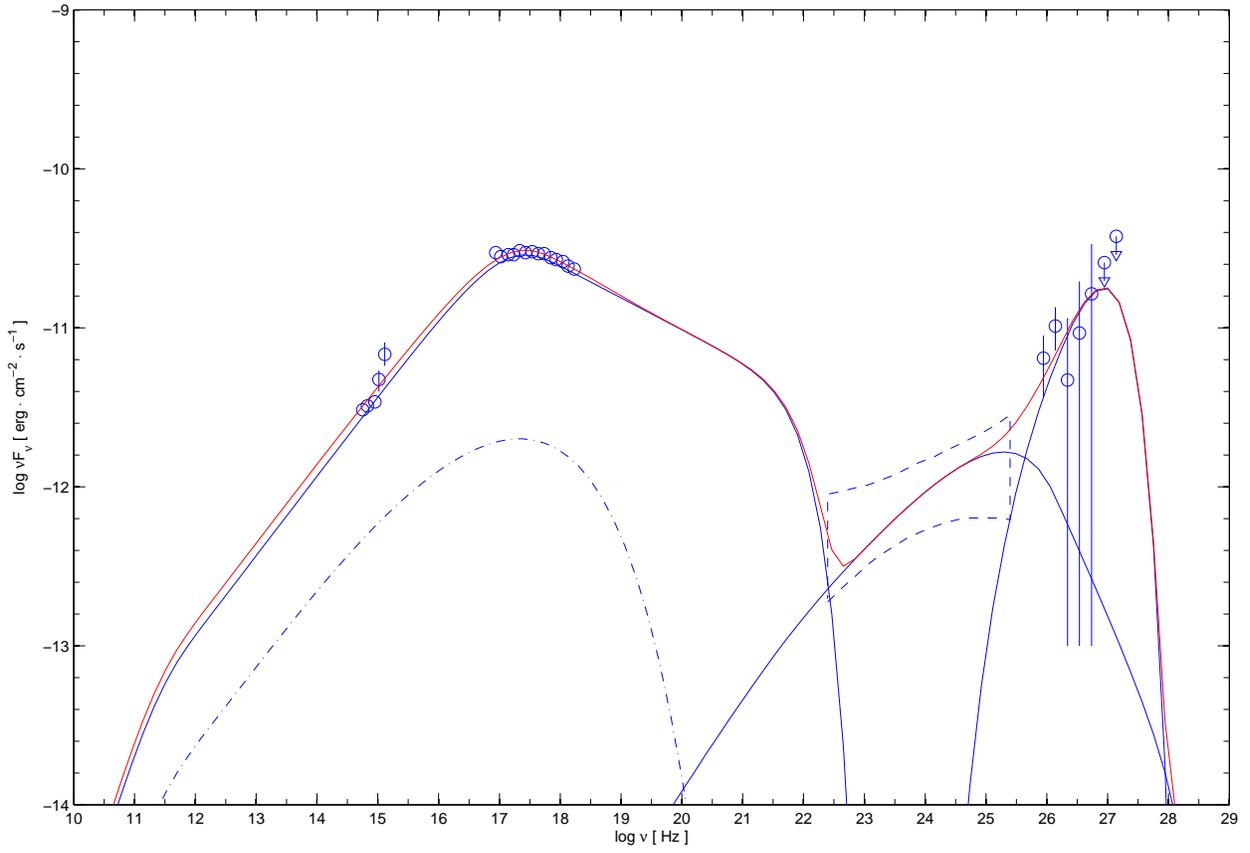}
\caption{Same as Fig. 1, but for the SED of 2004 June 5-10
\label{fig2}}
\end{figure}

\begin{figure}
\epsscale{1.}
\plotone{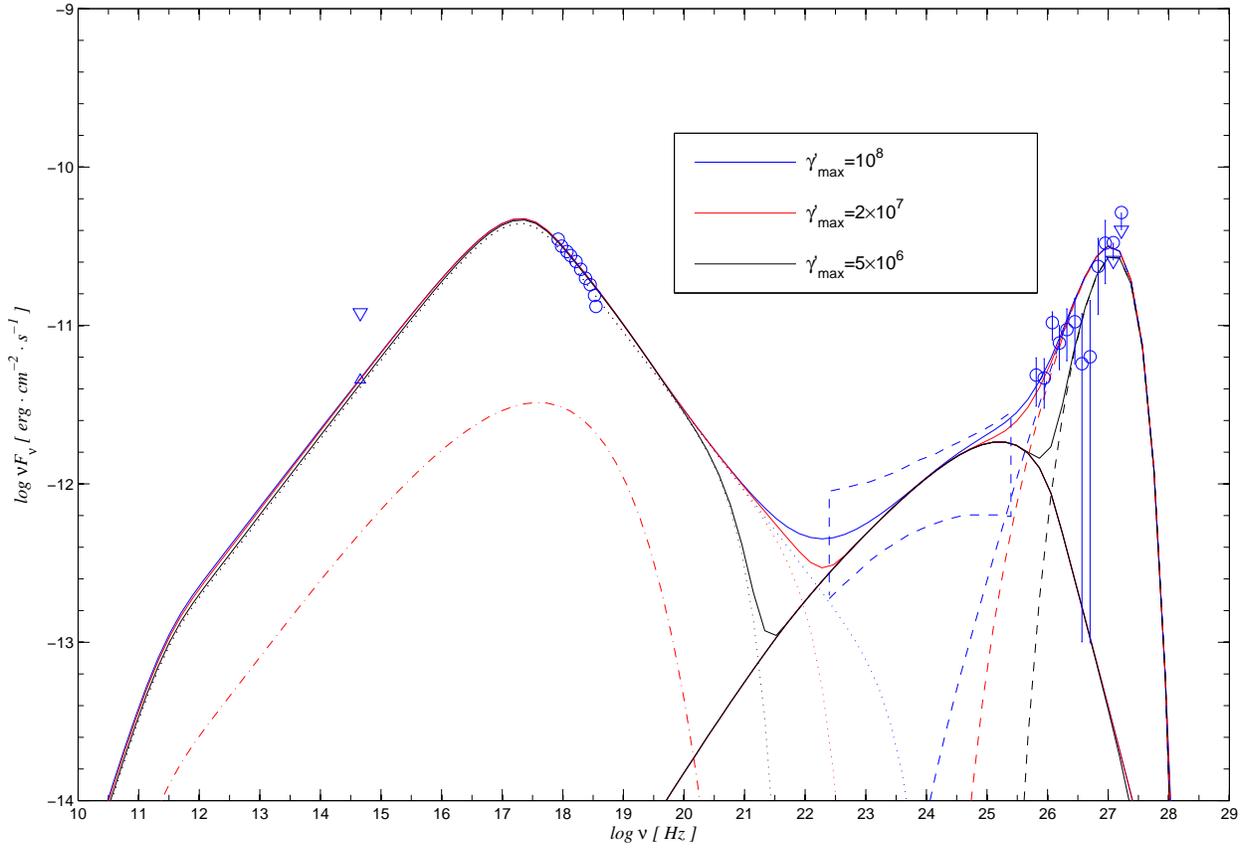}
\caption{Same as Fig. 1, but with various values of the maximum Lorentz factor $\gamma'_{\rm max}$.
$\gamma'_{\rm max}=5\times10^6$ (black), $\gamma'_{\rm max}=2\times10^7$ (red) and $\gamma'_{\rm max}=1\times10^8$ (blue)
\label{fig3}}
\end{figure}

\clearpage

\begin{table*}
\caption{Physical Parameters of the SSC Model} \label{para1}
\begin{center}
\begin{tabular}{lcccccccccccccccccccccccccccc}
\hline\hline
&Parameters & & & &2005 & & & &2004\\
\hline
&$\gamma'_{\rm min}$                      & & & &$1\times10^{2}$       & & & &$1\times10^{2}$\\
&$\gamma'_{\rm b}$                        & & & &$1\times10^{5}$       & & & &$9.5\times10^{4}$\\
&$\gamma'_{\rm max}$                      & & & &$2\times10^{7}$       & & & &$2\times10^{7}$\\
&$K_{\rm e}$ (cm$^{-3}$)                  & & & &54.1                  & & & &103.3\\
&$n_1$                                    & & & &2                     & & & &2\\
&$n_2$                                    & & & &4                     & & & &3.4\\
&$B$ (G)                                  & & & &0.4                   & & & &0.38\\
&$R'$ (cm)                                & & & &$4.0\times10^{16}$    & & & &$2.8\times10^{16}$\\
&$\delta_{\rm D}$                         & & & &15                    & & & &15\\
&$L_{\rm e}$ (erg  s$^{-1}$)              & & & &$1.9\times10^{43}$    & & & &$1.8\times10^{43}$\\
&$L_{\rm B}$ (erg  s$^{-1}$)              & & & &$2.2\times10^{44}$    & & & &$9.5\times10^{43}$\\
\hline
\end{tabular}
\end{center}
\end{table*}

\begin{table*}
\caption{Physical Parameters of the Proton Injection Spectrum} \label{para2}
\begin{center}
\begin{tabular}{lcccccccccccc}
\hline
&Parameters  & & & &2005 & & & &2004\\

\hline
&$E'_{\rm p,min}$ (eV)                   & & & &$1.45\times10^{8}$        & & & &$1.45\times10^{8}$ \\
&$E'_{\rm p,max}$ (eV)                   & & & &$1\times10^{14}$          & & & &$1\times10^{14}$\\
&$E'_{\rm c}$ (eV)                       & & & &$2.8\times10^{12}$        & & & &$2.8\times10^{12}$\\
&$N_{0}$ (eV$^{-1}$ cm$^{-3}$)           & & & &$5.0\times10^{19}$        & & & &$9.3\times10^{18}$\\
&$\alpha$                                & & & &2.2                       & & & &2.2\\
&$L_{\rm p}$ (erg  s$^{-1}$)             & & & &$2.7\times10^{53}$        & & & &$2.4\times10^{52}$\\

\hline
\end{tabular}
\end{center}
\end{table*}

\end{document}